\documentclass[12pt]{iopart}
\usepackage{graphicx}

\begin{document}

\title{Detection of basepair mismatches in DNA using graphene based nanopore device}

\author{Sourav Kundu}
\ead{sourav.kundu@saha.ac.in}
\address{Condensed Matter Physics Division, 
Saha Institute of Nuclear Physics, 
1/AF, Bidhannagar, Kolkata 700 064, India}

\author{S. N. Karmakar}
\ead{sachindranath.karmakar@saha.ac.in}
\address{Condensed Matter Physics Division, 
Saha Institute of Nuclear Physics, 
1/AF, Bidhannagar, Kolkata 700 064, India}
 
\begin{abstract} 
  We present a new way to detect basepair mismatches in DNA leading to 
different epigenetic disorder by the method of nanopore sequencing. Based on 
a tight-binding formulation of graphene based nanopore device, using Green's 
function approach we study the changes in the electronic transport properties 
of the device as we translocate a double-stranded DNA through the nanopore embedded 
in a zigzag graphene nanoribbon. In the present work we are not only successful 
to detect the usual AT and GC pairs, but also a set of possible mismatches in the 
complementary base-pairing. 
  
\end{abstract} 

\pacs{72.15.Rn, 73.23.-b, 73.63.-b, 87.14.gk} 

\noindent{\it Keywords\/}: DNA sequencing, Graphene nanopore, Basepair mismatch.

\submitto{\NT}

\maketitle

\section{Introduction:}

 Basepair mismatches in DNA is one of the major reason behind several mutagenic 
disorders which may lead to different genomic instabilities, development of cancer~\cite{loft} 
and other degenerative diseases. Mismatch in DNA bases occurs mainly due to 
misincorporation of nitrogen bases during DNA replication, oxidative or chemical 
damages and ionizing radiations. Inspite of dramatic advancements in medical 
science, many crucial issues, such that how DNA detect and repair damages, 
individual mismatches or what is the most accurate observable physical parameter 
to detect basepair mismatch is still remain clouded. Apart from traditional 
fluorescence-based sequencing technique~\cite{sanger}, several other methods 
also applied to detect mismatches. Some examples are magnetic signatures~\cite{tapash1}, 
longitudinal electronic transport~\cite{tapash2,tapash3} thermodynamic properties 
of basepair mismatches~\cite{tapash4} and study of stretched DNA using AFM~\cite{zhang}, 
but no conclusive results appear. Whereas with the advent of nanopore-based 
sequencing~\cite{bayley,branton,kasiano,deamer,lagerq,sigalov} 
a new pathway is opened for marker-free gene testing. In early days of nanopore 
sequencing people mostly used biological nanopores ($\alpha$-Haemolysin), detect 
the changes in ionic current as a single-stranded DNA (ss-DNA) passes through the 
pore~\cite{bayley,branton,kasiano,deamer}. With time, usage of nanopore materials 
also evolves from biological to solid state nanopores. The latter one overcomes many 
drawbacks of biological nanopores {\it e.g.}, poor mechanical strength~\cite{striemer}, 
problems of integration with on-chip electronics~\cite{rosen}. Solid state nanopores 
also provides some other advantages like multiplex detection~\cite{kim} and different 
detectable physical parameters other than ionic current~\cite{lagerq,zwolak,gracheva,
sigalov,mcnally,huang,tsutsui,xie}. Though it provides so many advantages but it lacks 
in an important case, the average thickness of synthetic membranes used for molecular 
detection is of the order of 10 nm, which will occupy several nucleobases at a time 
(distance between two consecutive nitrogen bases in a DNA chain is 0.34 nm), 
jeopardizing single molecule base-specific detection. Graphene, single layer of 
graphite~\cite{novo1}, provides a solution to this problem. As the single layer thickness 
is of the order of the distance between two consecutive bases in DNA and with various 
advantageous properties~\cite{rocha} it is the ideal candidate for sequencing applications 
(recently other 2-D material, such as silicene also has been studied for the purpose of DNA 
sequencing~\cite{ralph1}). Graphene also provides several ways of sequential detection {\it e.g.}, 
nanoribbon conductance~\cite{nelson,min,saha}, transverse tunnelling~\cite{postma,prasong}. 
Readers can consult some review articles~\cite{fyta,ralph,rashid,ventra1} for a detailed 
description of nanopore based sequencing techniques.

 In this work we present a theoretical study to detect basepair mismatches 
in DNA using the method of nanopore sequencing. Though several studies on 
ss-DNA sequencing already exists in literature~\cite{postma,saha,nelson,he,pathak}, 
there is no such report on double-stranded DNA (ds-DNA). We use a graphene nanopore 
based sequencing device which is created on single layer zigzag graphene nanoribbon 
(zgnr) following Ref.~\cite{saha}. Using Landauer-B{\"u}ttiker formalism we study the 
changes in electronic transport properties of the device as a ds-DNA (which also contains 
basepair mismatches) translocates through the nanopore. Distinct features have been 
observed in transmission probability and to some extent in I-V response also for the 
canonical Watson-Crick pairs and for four different types of possible mismatches. 
Study of local density states (LDOS) also provide applicable insight. Our results open 
a new pathway for reliable detection of basepair mismatches in DNA, a highly important 
diagnosis for genetic disorder.

\section{Theoretical Formulation:} 

\begin{figure}[ht]
  \centering

    \includegraphics[width=65mm,height=50mm]{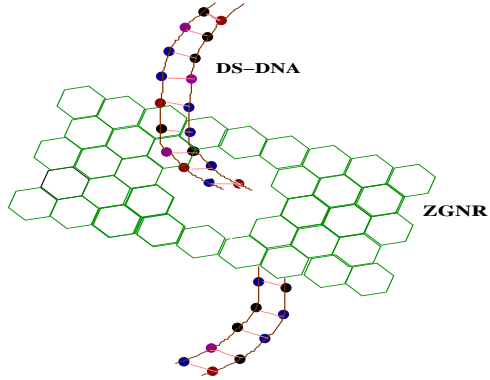}

\caption{(Color online). Schematic view of the ZGNR nanopore device with 
a ds-DNA passing through the nanopore. Current is lateral through the zgnr 
{\it i.e.}, in the trnasverse direction.}

\label{fig1}
\end{figure}

 To perform numerical study on the sequential determination of basepair 
mismatches in DNA we use zigzag graphene nanoribbon, with a pore created 
at the centre of it. We preserve the two-sublattice symmetry of graphene 
while creating the nanopore~\cite{sourav}. The whole zgnr system can be 
presented by an effective Hamiltonian (see Fig.~\ref{fig1}) 

\begin{eqnarray}
& H_{zgnr}&= \sum\limits_{i=1}^N\left(\epsilon
c^\dagger_{i}c_{i}+t c^\dagger_{i}c_{i+1}+\mbox{H.c.} \right)
\end{eqnarray}

 where $\epsilon$ is the site-energy of each carbon atom in ZGNR, 
and $t$ is the nearest neighbour hopping amplitude. $c_i$ and 
$c^\dagger_{i}$ creates or annihilates an electron at the ith site 
respectively. For calculation of transport properties we also use 
semi-infinite zgnr as electrodes~\cite{saha}. Thus the total Hamiltonian 
of the system can be written as $ H_{tot}=H_{zgnr}+ H_{leads}+H_{tun}~,$
where $H_{tun}$ 
represents tunneling Hamiltonian between the nanopore device and electrodes. 
In our calculations we scale energy in terms of t {\it i.e.}, 
we set t=1.0 eV. 
Hamiltonian of a ds-DNA can be expressed as 

\begin{eqnarray}
& H_{DNA}&= \sum\limits_{i=1}^N\sum\limits_{j=I,II}\left(\epsilon_{ij}
c^\dagger_{ij}c_{ij}
+t_{ij}c^\dagger_{ij}c_{i+1j}+\mbox{H.c.} \right)\nonumber \\
&&~~~~~~~~~~~~~~~~~+ \sum_{i=1}^N v \left(c^\dagger_{iI}c_{i II}+
\mbox{H.c.} \right)~
\end{eqnarray} 
where $c_{ij}^\dagger$ and $c_{ij}$ are the electron creation and annihilation 
operators at the {\it i}th nucleotide of the jth strand, $t_{ij}=$ nearest 
neighbour hopping amplitude between nucleotides along the jth strand, 
$\epsilon_{ij}=$ on-site energy of the nucleotides, $v=$ interstrand 
hopping between the nucleobases.
 
 Green's function formalism is used for both the LDOS and transport 
properties calculations. Transmission probability of an electron with an 
energy E is given by $T(E)={\mbox {\rm Tr}} [\Gamma_L G^r \Gamma_R G^a]$~\cite{datta1}, 
where $G^r=[G^a]^\dagger$ and $\Gamma_{L(R)}=i[\Sigma^r_{L(R)}-\Sigma^a_{L(R)}]$. 
$G^r=[E- H_{zgnr}-\Sigma^r_L-\Sigma^r_R+i\eta]^{-1}$ is the single-particle 
retarded Green's function for the entire system at an energy E, where 
$\Sigma^{r(a)}_{L(R)}=H^\dagger_{\mbox{tun}} G^{r(a)}_{L(R)} H_{\mbox{tun}}$ 
represents retarded (advanced) self energies of the left (right) zgnr electrodes which 
is calculated following recursive Green's function technique~\cite{nardelli, lopez}. 
$G^{r(a)}_{L(R)}$ is the retarded (advanced) Green's function of the left (right) lead. 
At absolute zero temperature, using Landauer formula, current through the 
nanopore device for an applied voltage V is given by 
$I(V)=\frac{2e}{h} \int^{E_F+eV/2}_{E_F-eV/2} T(E)dE~$ 
where $E_F$ being the Fermi energy. Here we assume that there is no charge 
accumulation within the system. The LDOS profiles of the basepairs 
trapped inside the nanopore are given by 
 $\rho(E,i) = - \frac{1}{\pi} {\rm Im[G_{ii}(E)]}$
where, $G(E)= (E-H+i\eta)^{-1}$ is the Green's function for the zgnr system 
including the basepairs with electron energy E as $\eta\rightarrow0^+$, $H=$ 
Hamiltonian of the zgnr-nanopore, and, ${\rm Im}$ represents imaginary part of 
$G_{ii}(E)$. $G_{ii}(E)$ is the diagonal matrix element $(< i|G(E)|i >)$ of the 
Green’s function, $|i >$ being the Wannier state associated with the trapped 
nucleotide.

\section{Results:}
 For the purpose of numerical investigation we use ionization potentials of the 
nitrogen bases as their site energies which are extracted from the {\it ab-initio} 
calculations~\cite{senth}: $\epsilon_G$= 8.178, $\epsilon_A$= 8.631, 
$\epsilon_C$= 9.722,and $\epsilon_T$= 9.464, all units are in eV. Then we shift 
the reference point of energy to the average of the ionization potentials of the 
nucleobases  which is 8.995 eV, and with respect to this new origin of energy 
the on-site energies for the bases G, A, C, and T become -0.82 eV, -0.37 eV, 
0.72 eV, and 0.47 eV respectively. This is valid for model calculations as it 
won't do any qualitative damage to the results. Similar methods have previously 
been employed where the average of ionization potential is set as the backbone 
site-energy~\cite{paez}.  

\begin{figure}[ht]
  \centering

    \includegraphics[width=65mm,height=45mm]{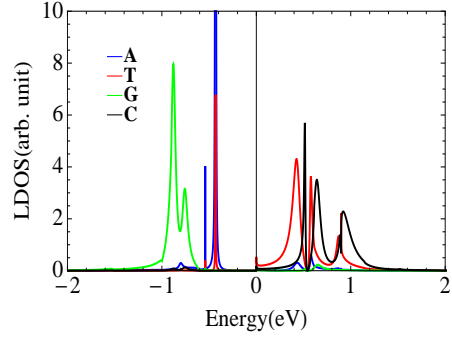}

\caption{(Color online). LDOS of the four nucleotides trapped at 
the nanopore. There are four distinct peaks of different heights 
representing different bases close to their characteristic site-energy.}

\label{fig2}
\end{figure}

 In Fig.~\ref{fig2} we show the LDOS profiles for the four different nitrogen 
bases. We study this LDOS response of the bases as a part of the Watson-Crick 
basepairs not as individual {\it i.e.}, we trap the AT and GC pairs inside the 
nanopore and study the LDOS profile of the respective bases. The position of 
different peaks in the LDOS are different, close to the characteristic site 
energies of the different nucleotides and the peak values are also different. 
These relative differences in LDOS patterns present a chance to detect the basepairs 
using ARPES technique by trapping them inside the nanopore. As the LDOS behaviour 
is mostly dominated by the nitrogen bases not by the backbones~\cite{he} this also 
provides a new way of biomolecular detection.

\begin{figure}[ht]
  \centering
  \begin{tabular}{cc}

    \includegraphics[width=45mm,height=34mm]{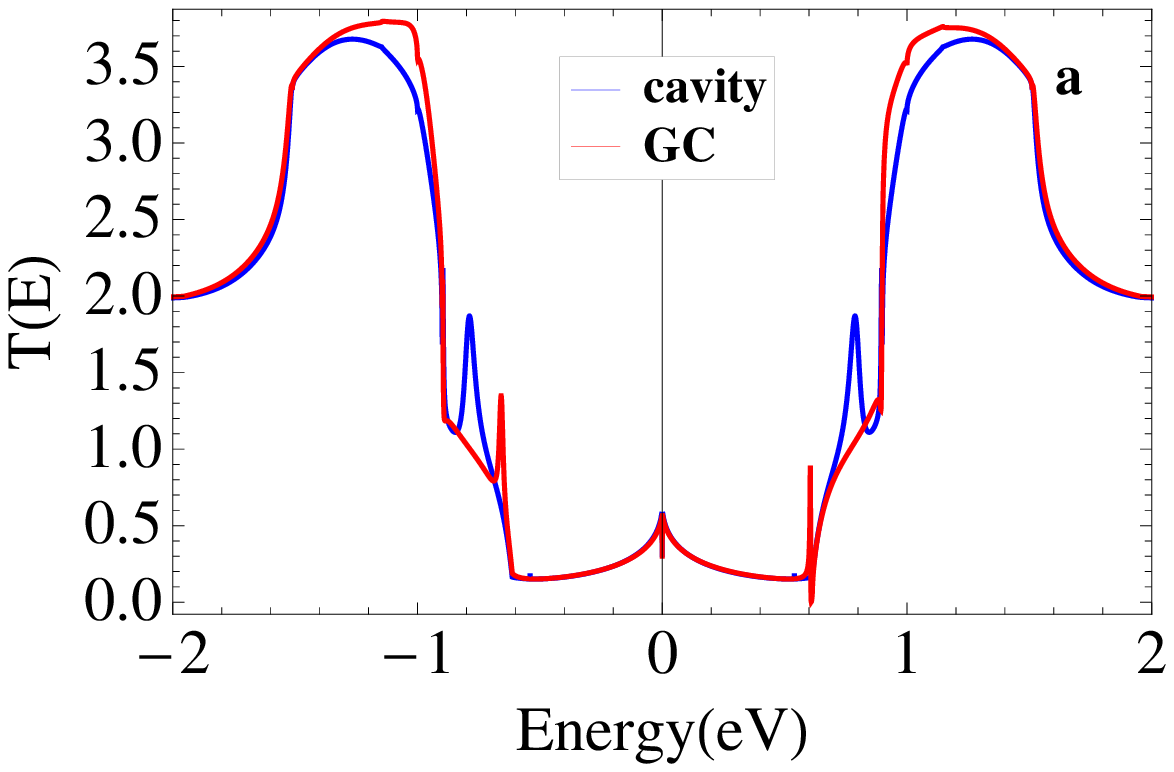}&
   
    \includegraphics[width=45mm,height=34mm]{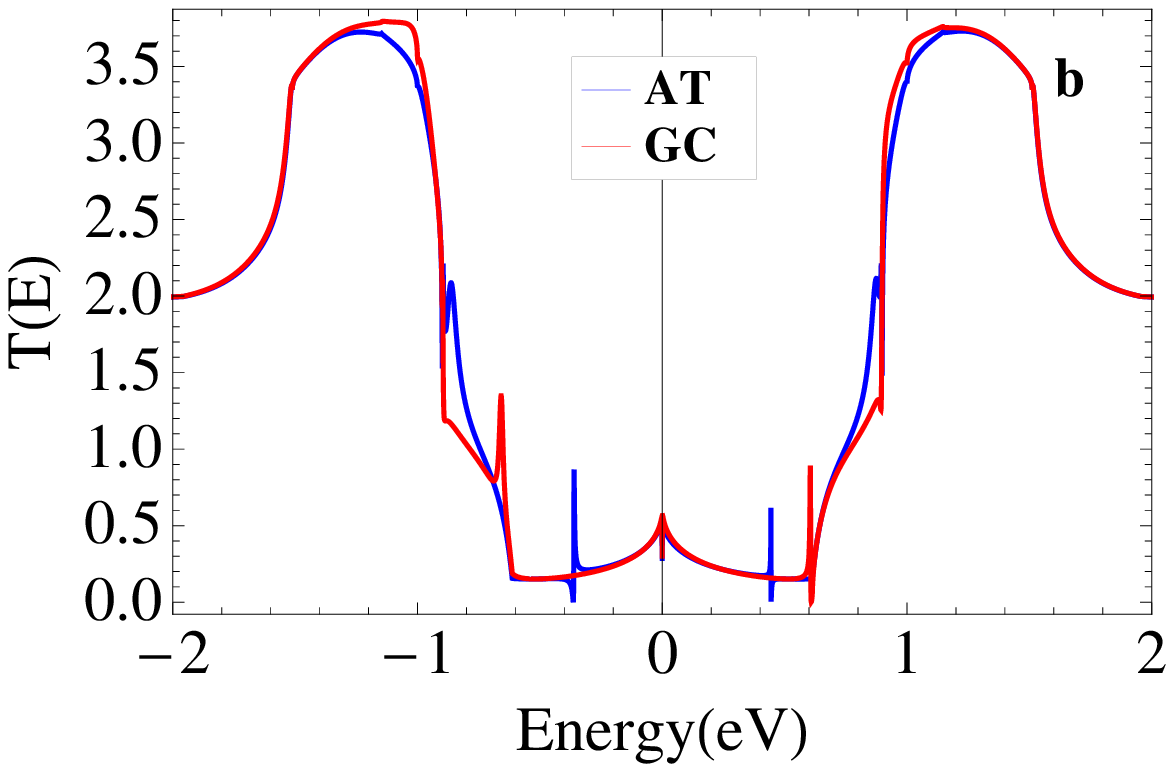}\\

    \includegraphics[width=45mm,height=34mm]{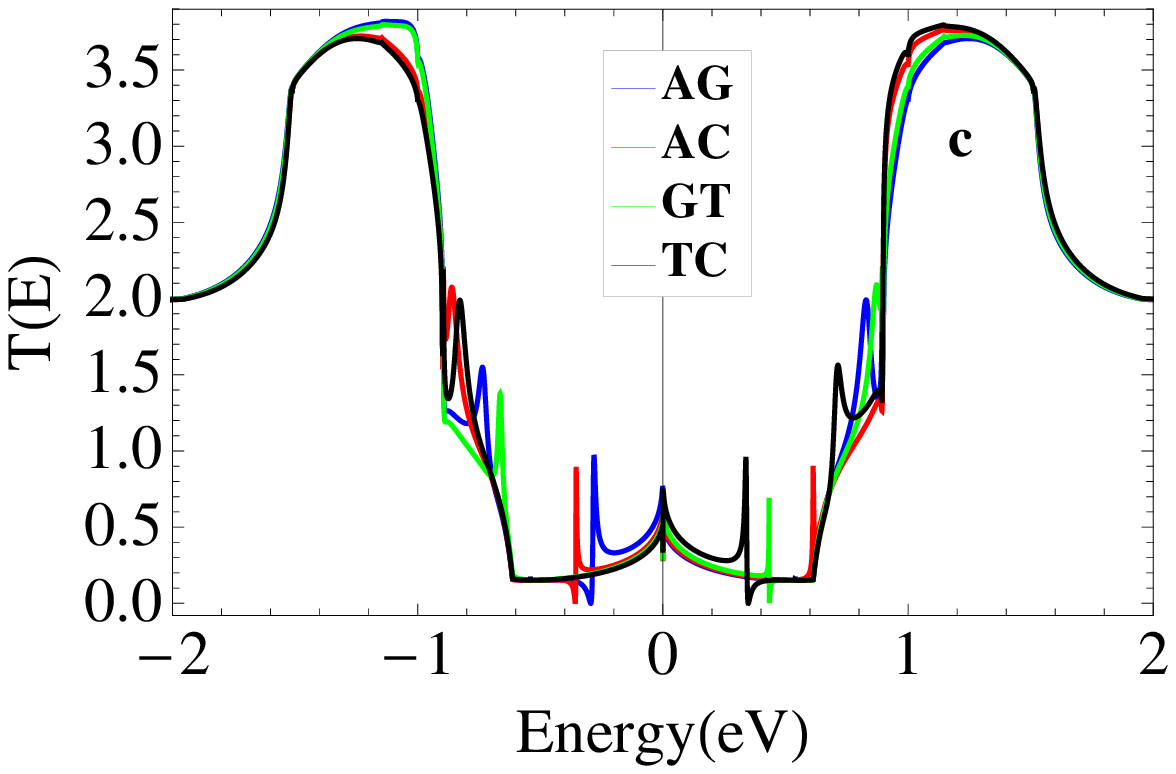}&

    \includegraphics[width=45mm,height=34mm]{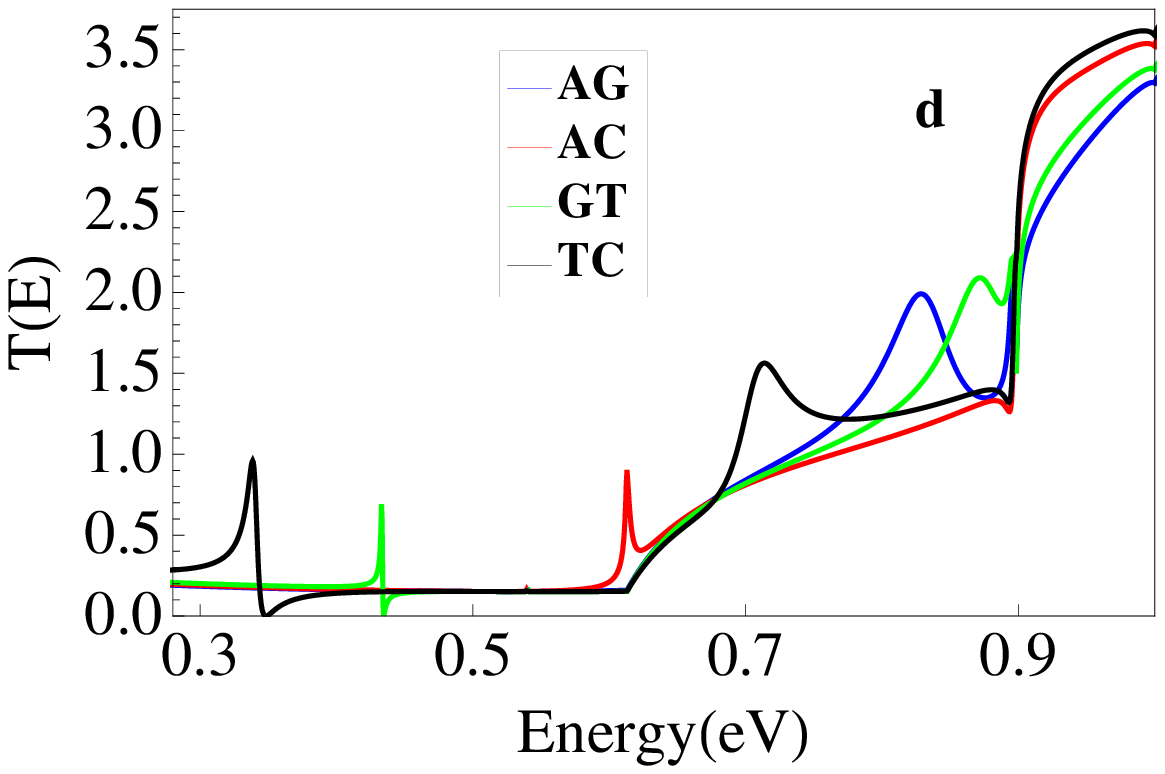}\\

 \end{tabular}
\caption{(Color online). Transmission probability T(E) as a function of energy for different cases. 
a) Comparison between a bare nanopore and GC-nanopore. b) Comparison between two Watson-Crick pairs 
AT and GC. c) Characteristic features of four different mismatched basepairs trapped inside the 
nanopore. d) Enlarged view of the plot (c) for clear visualization.}
\label{fig3}
\end{figure}

 In Fig.~\ref{fig3} we plot the variation in transmission probability for different 
cases. The coupling parameter between the boundary sites of the zgnr-nanopore and 
DNA base is set to 0.2 eV. Intrastrand hopping parameter between identical bases in the 
DNA chain is taken as $t_{ij}$=0.35 eV and for different bases $t_{ij}$=0.17 eV. Whereas 
interstrand hopping between nucleobases is taken as v=0.035 eV, one order of magnitude less 
than the intrastrand hopping. These values are consistent with previous reports~\cite{paez,
klosta,sourav1,sourav2,sourav3}. Fig.~[3a] shows the comparison between a bare nanopore 
and a DNA basepair ( GC pair ) trapped into the nanopore. The changes in transmission 
spectra are clearly distinguishable. There are characteristic peaks in the profile 
both at the +ve and -ve energy range. Both the curves for bare nanopore and GC-nanopore 
are symmetric with respect to zero of energy, as the two-sublattice symmetry of the 
graphene nanopore is preserved in both the cases. It was violated in case of 
ss-DNA sequencing~\cite{sourav}. Fig.~[3b] shows the difference between the 
characteristic features of two Watson-Crick pairs AT and GC. Distinct peaks are present 
in the transmission profile at and around the characteristic site energies of the 
respective nucleobases. In Fig.~[3c] we show the relative changes in the transmission 
profile for four different types of basepair mismatches. Each of the mismatches has 
distinct response at and around their respective site energies. Variations are quite 
similar in +ve and -ve energy range. They are clearly distinguishable at low energy, 
and the characteristic features die down as we move towards higher energy values. It is 
due to the fact that as we go to higher energy we are moving away from the characteristic 
site energies of the nucleobases. In Fig.~[3d] we zoom in a small energy window of 
Fig.~[3c] for better visualization. TC mismatch has distinct peak around 0.3 eV. 
GT and AC mismatches become clearly distinguishable between 0.4 to 0.45 eV and 0.6 
to 0.65 eV respectively. Whereas AG becomes visibly distinct in the energy range 
0.8 to 0.9 eV. 

\begin{figure}[ht]
  \centering
  \begin{tabular}{cc}

    \includegraphics[width=45mm,height=34mm]{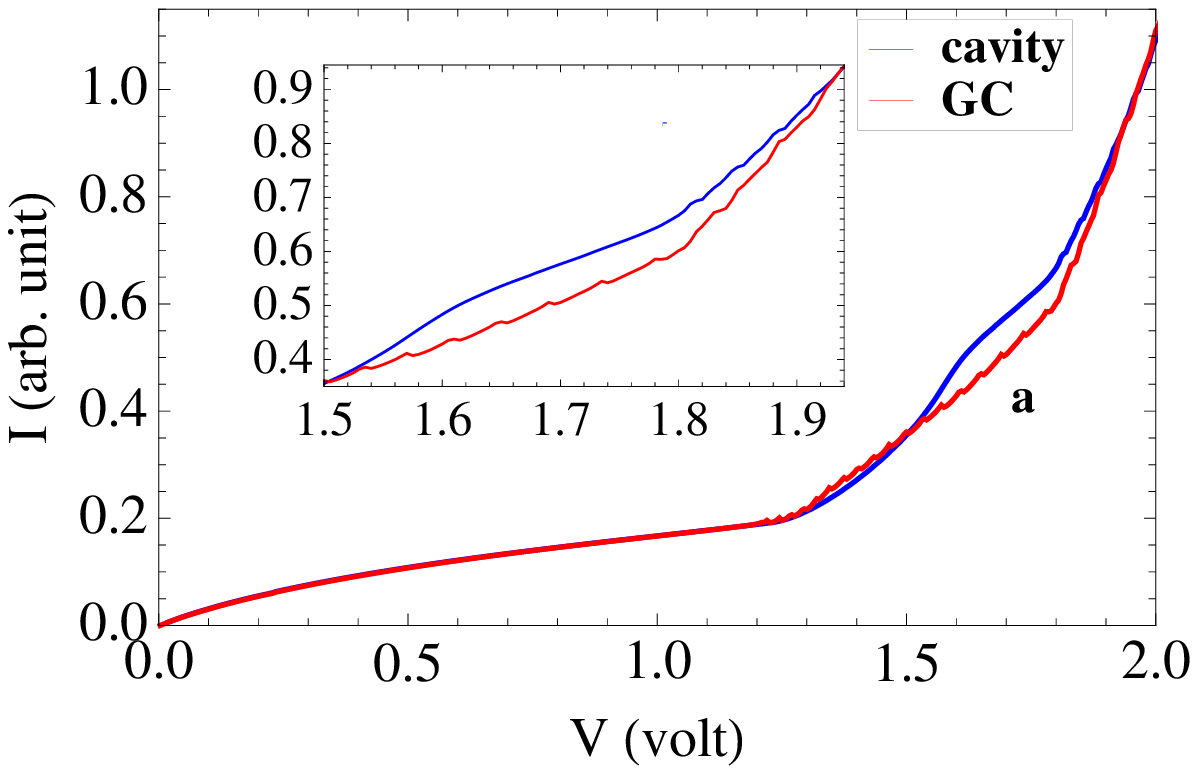}
   
    \includegraphics[width=45mm,height=33.65mm]{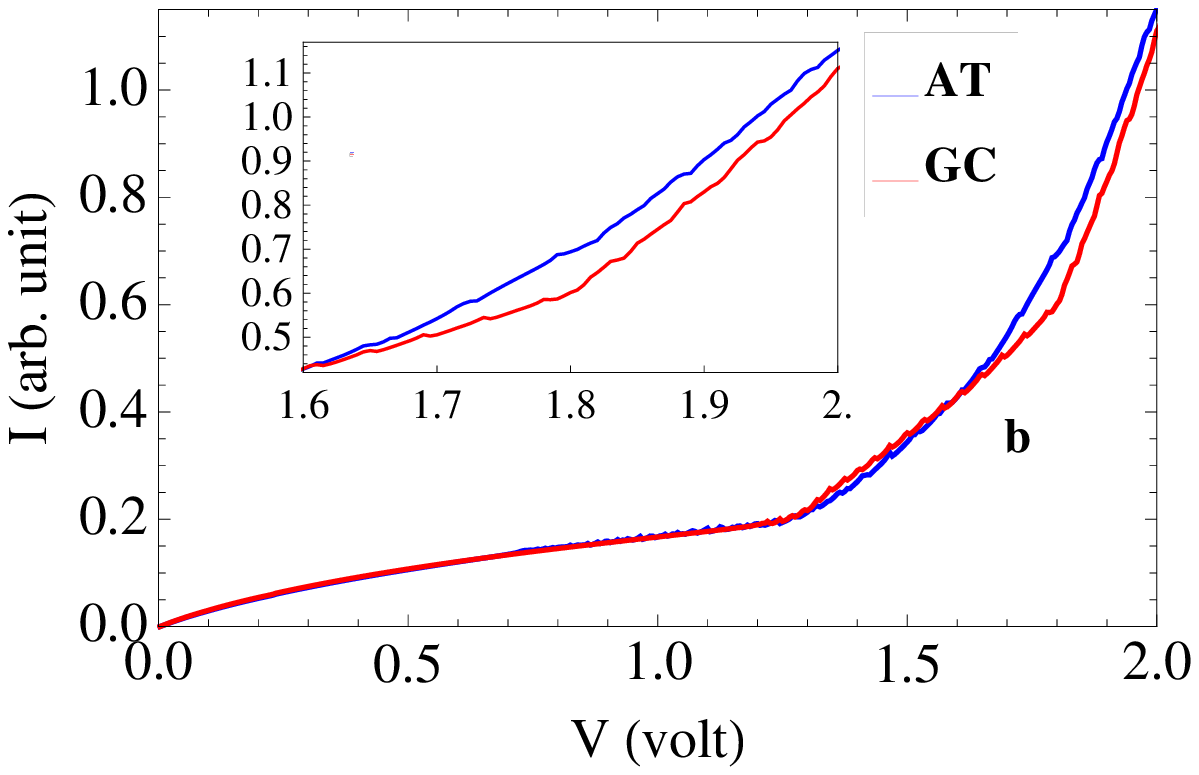}\\
    
    \includegraphics[width=60mm,height=38mm]{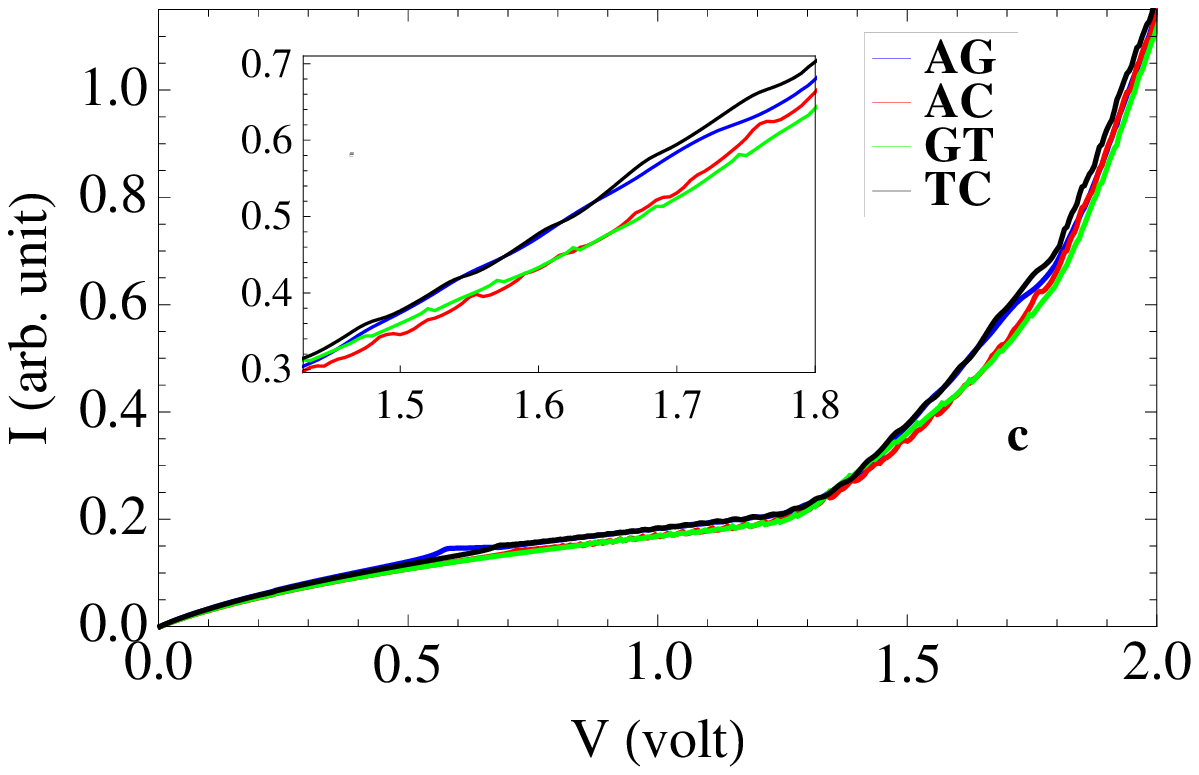}

 \end{tabular}
\caption{(Color online). Current - Voltage response of the active nanopore 
device for different cases. a) Comparison of the current responses between a 
bare nanopore and GC-nanopore. b) Difference between characteristic current 
amplitudes of two Watson-Crick basepairs AT and GC. c) Attributes of four 
different mismatches (AG, AC, GT, TC). Insets show selective voltage ranges 
for better visualization. $E_F$=0 eV represents Fermi energy.}
\label{fig4}
\end{figure}

 In Fig.~[4a] we show changes in the I-V characteristics for a bare nanopore 
and a GC-nanopore. Effect of the basepair inside the nanopore becomes 
prominent at considerable bias, inset shows a specific high voltage range of 
the curves where they are clearly distinguishable. Fig.~[4b] shows the variation 
in the current response between two Watson-Crick pairs AT and GC. They also 
become differentiable at high voltage range between 1.7 to 2.0 Volt. AT pair 
produces higher current than GC pair, which reflects their different electronic 
structure, as this current response depends on how the local charge density 
profile modified due to the insertion of the DNA bases~\cite{nelson}. Fig.~[4c] 
shows the relative differences between four possible mismatches of basepairs. At 
low bias differences between them is very faint, they gradually become differentiable 
as we increase the bias. Insets of the Fig.~[4b] and Fig.~[4c] show specific voltage 
windows within which the mutual separation between the basepairs is larger than elsewhere.

\begin{figure}[ht]
  \centering
  \begin{tabular}{cc}

    \includegraphics[width=49.5mm,height=40mm]{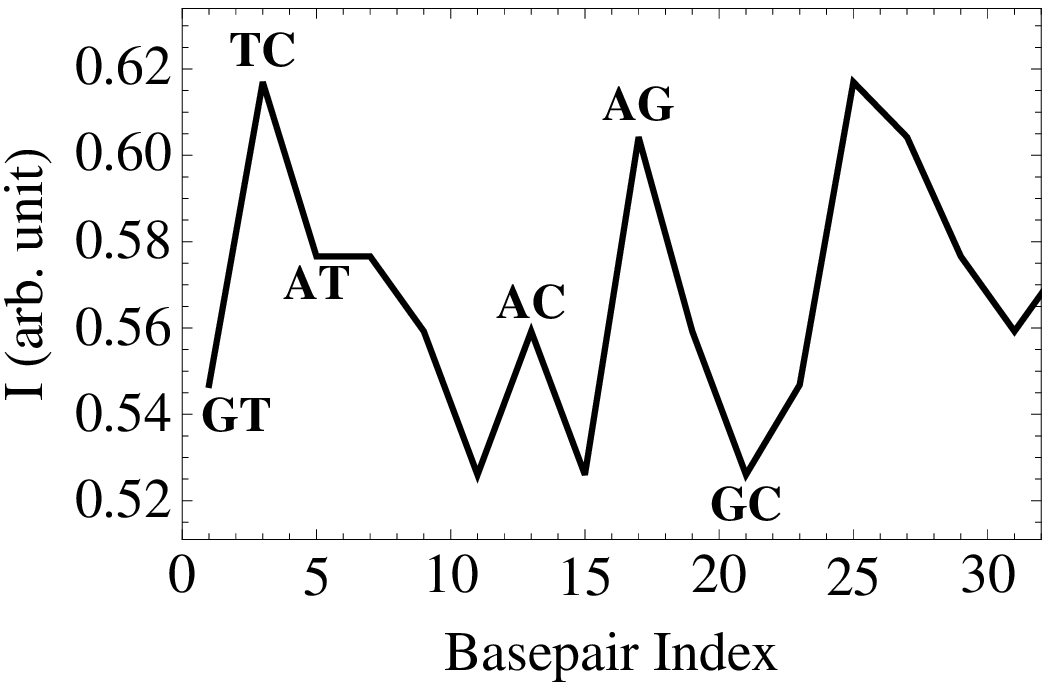}
   
    \includegraphics[width=49.5mm,height=40mm]{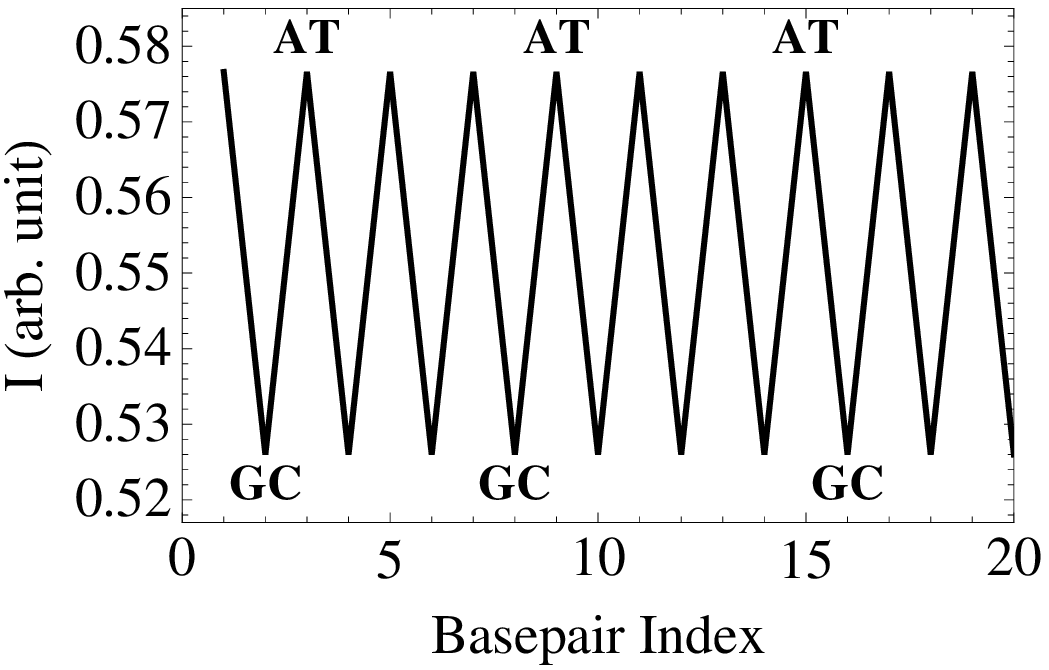}
\end{tabular}
\caption{(Color online). Left panel shows the stop and go translocation of a 
Random ATGC ds-DNA chain through the nanopore, while bias across the device is 
fixed to a specific value which gives maximum separation in current response for 
different basepairs. We record the characteristic current output for the bases 
as they translocate through the nanopore. The respective basepairs and mismatches 
are indicated in the figure with their usual symbols (AT, GC etc.). Right panel 
shows the same variation for a ds-DNA chain with no basepair mismatch for better 
understanding of the left panel figure. Though the current is presented in arb. 
unit as we report a model calculation, but if we put the exact numerical values of 
different constants like h, e and $\hbar$, it turns out of the order of 10 $\mu$A.} 

\label{fig5}
\end{figure}

 In Fig.~\ref{fig5} we finally show the sequencing application to detect basepair 
mismatches along with the two canonical pairs AT and GC. We take a 30-basepair 
long Random ATGC chain, translocate it through the nanopore and record the 
characteristic current signals corresponding to the different basepairs including 
the mismatches. During this translocation bias is kept at 1.72 Volt, this voltage 
gives maximum possible relative separation between the characteristic currents 
of different basepairs (see insets of Fig.[4b] and Fig.[4c]). Separation between 
a Canonical pair GC and a mismatch TC is maximum whereas that between AT and AC 
is minimum. The reason behind this is G and T are from different group, G is from 
purine group and T is from pyrimidine, electronic structure of them are also quite 
different. So when the pairing changes from GC to TC, the corresponding change in 
current response is also big. While for AT and AC, both T and C are from the same 
pyrimidine group, hence the relative changes in the response is also quite smaller. 
These relative changes in the current response represent the difference in their 
electronic structure. If we define a new quantity to measure the sensitivity 
of this type of sequencing devices {\it e.g.}, percentage separation =$(I_{max}-I_{min})/I_{min}$, 
it turns out to be that maximum and minimum values of percentage separation achieved 
are 17.30$\%$ and 3.23$\%$ which implies that the current signals for the respective 
basepairs can be detected with much more reliability. We also plot a separate 
figure (see right panel of Fig.~[5]) for a normal ds-DNA chain without any mismatches, 
for better understanding of the effect of mismatches on the current response of the device. 
It is also important to mention that though we have presented current in arbitrary unit, 
but if we put numerical values of various constants {\it e.g.}, h, e and $\hbar$, it turns 
out that the currents are of the order of 10 $\mu$A which is much higher than previous 
reports on ss-DNA sequencing as well as much greater than the noise level of this type 
of devices which is of the order of nA~\cite{saha}. Very recently a report by Feliciano 
{\it et al.}~\cite{ralph} on dynamical effects of environment on operation of graphene based 
sequencing devices shows that fluctuations of the nucleotides inside the nanopore may change 
the conductance of the devices relying on tunneling mechanism, though they conclude that these 
effects would not be very important for the devices which relies on transverse conductance with 
larger transmission probability. As our proposed device relies on transverse conductance and 
produces greater current output, effect of these type of noises will be much lesser.
Whereas another study by Krems {\it et al.}~\cite{krems} in 2009 dealing with 
different types of noises which may occur in actual sequencing experiments showed that these 
environmental effects do not strongly influence the current distributions and working efficiency 
of these devices. Though based upon these results we can say that the overall sensitivity of our 
device won't be hampered too much but there will always be sources of noise in actual experimental 
condition due to environmental fluctuations, presence of water and counterions which can affect 
the device operation. It is also important to note that it is one of the early attempt to detect 
basepair mismatches by means of nanopore sequencing and the results given in this work is open to 
improvement in different ways. One example is, by functionalization of the edge atoms of the nanopore 
which can significantly enhance nucleobase-pore interaction, thus reducing the structural noise by 
enhancing the graphene-nucleobase electronic coupling~\cite{he1,garaj}. Different types of groups 
can be used for functionalization (e.g., hydroxide~\cite{jeong}, amine or nitrogen~\cite{saha}) to 
provide custom made solution to overcome noise in electrical DNA sequencing techniques. It is also 
true for the devices relying on transverse conductance that most of the current passes through the 
edges of the nanoribbon which is one of the reason of poor sensitivity of these type of devices,
but this can be controlled with accurate engineering of the nanopore device dimension. See Appendix 
section for more details on this.

\section{Conclusion:} 
 
 In summary we present an effective and reliable technique to detect basepair 
mismatches in a given DNA sample. We analyze different properties from LDOS to 
I-V response in connection with sequential determination and found distinguishable 
signatures in most of the cases. Most of the earlier results on DNA sequencing 
use ss-DNA which neglect the basic problem of basepair mismatch leading to different 
neuro-degenerative diseases. As the different genetic diseases occur due to mismatch 
of base-pair {\it i.e.}, when a nitrogen base in a DNA double-helix paired up with 
another base which is not the complementary pair of it, sequencing of ss-DNA can't 
provide this information. On the other hand previous attempts to detect basepair 
mismatches do not provide any decisive results. With time both medical science 
and genetic research progress, the reasons behind different genetic disorder 
including neuro-degenerative ones (like Perkinsons, Alzheimer etc) are becoming 
more and more transparent. With this progress the need for low cost and reliable 
DNA sequencing also increases which should also provide the necessary technique 
for proper medical applications. In this circumstances we present a reliable 
tight-binding scheme to detect basepair mismatches in DNA with much better 
accuracy than previous studies~\cite{tapash5}. At the same time, we also 
understand that proposed technique needs more improvements for actual application 
in real environment and hope it will soon be tested with further modifications.

\section{Appendix:}

In this section we provide some additional information on basepair detection of DNA. 
In Fig.~[6] we plot the variation in the current response of our proposed 
device for AT and TA basepairs, both are being Watson-Crick pair. Now for the previous 
calculation we preserve the two-sublattice symmetry of graphene by symmetrically 
connecting the nucleotides with edge atoms, in this configuration it is hard to distinguish 
AT and TA separately. For better detectability we destroy the two-sublattice symmetry and 
find distinct responses. The same also has been done for detection between GC and CG. We 
want to mention that we checked all our results with broken sublattice symmetry, but find 
no significant changes for the results presented in the earlier sections. The percentage 
seaparation between AT and TA (GC and CG) is relatively small (1.5$\%$) which implies 
that the proposed device is not effective in the same way as it is for basepair mismatches.

\begin{figure}[ht]
  \centering
  \begin{tabular}{cc}

    \includegraphics[width=47mm,height=40mm]{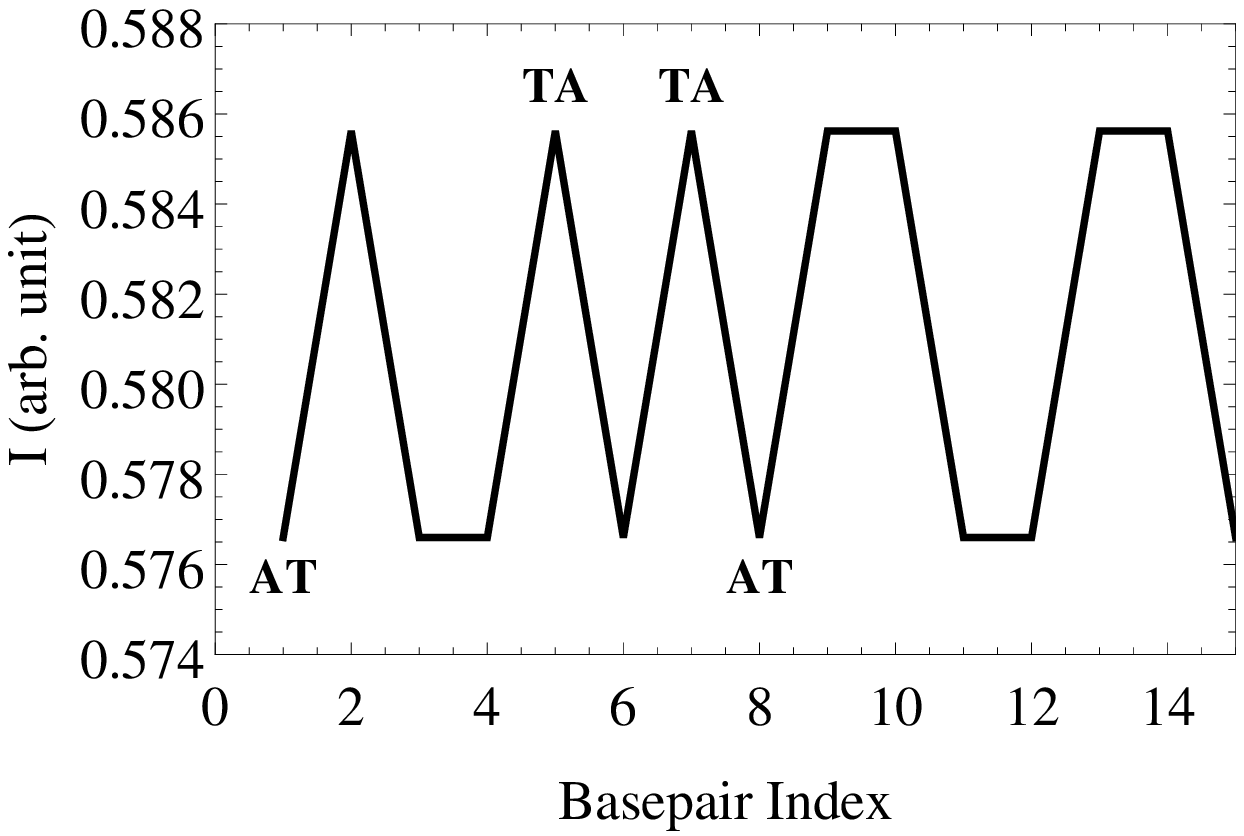}
   
    \includegraphics[width=47mm,height=38.5mm]{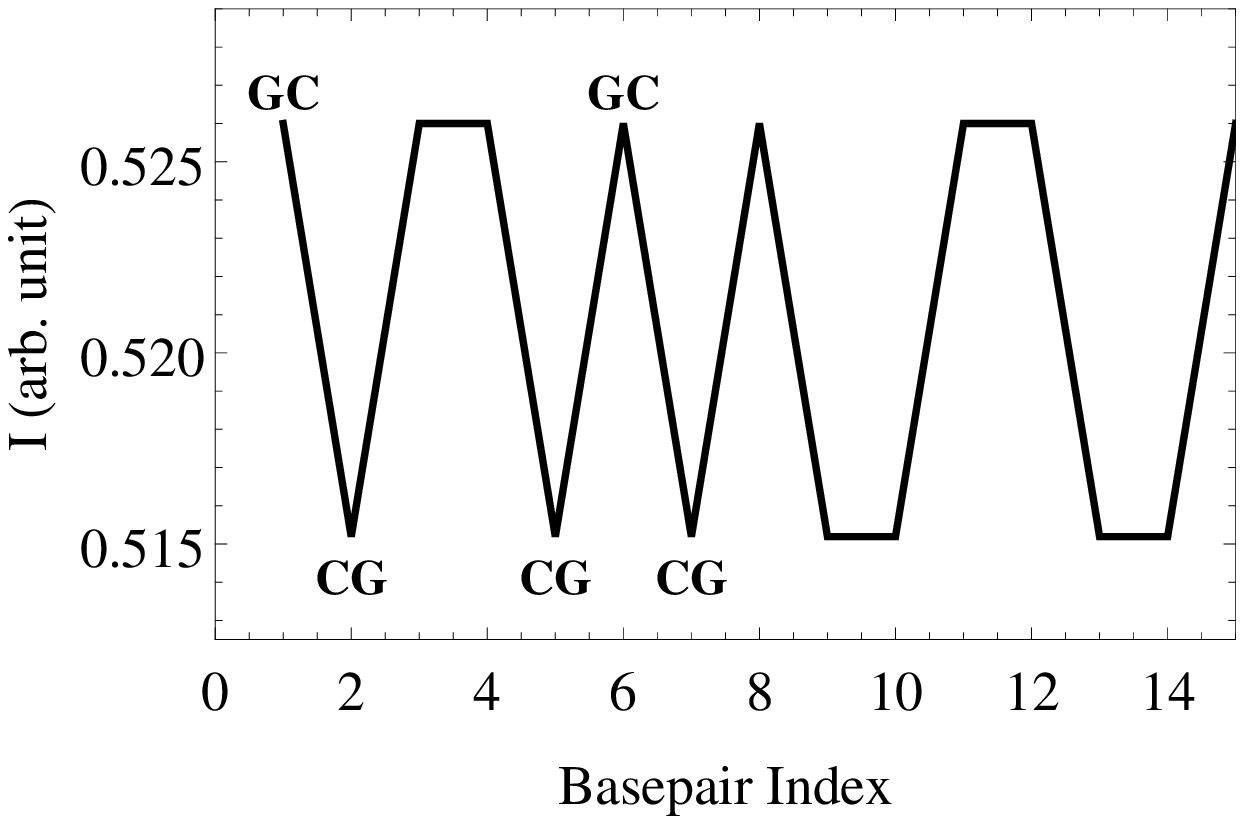}\\

 \end{tabular}
\caption{(Color online). Current - Voltage response of the active nanopore 
device for two Watson-Crick pairs in opposite orientation. Figure on the left 
side shows comparison of the current responses between a AT-nanopore and 
TA-nanopore. Right panel shows the same for GC-nanopore and CG-nanopore.}
\label{fig6}
\end{figure}

\begin{figure}[ht]
  \centering

    \includegraphics[width=63mm,height=45mm]{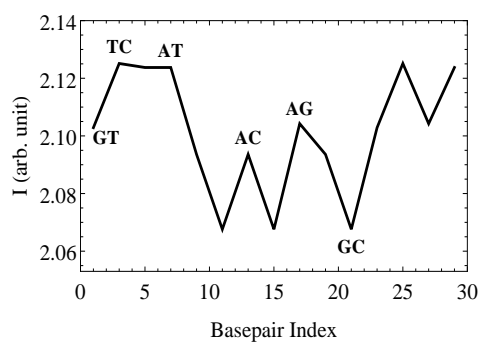}

\caption{(Color online). Stop and go translocation of a Random ATGC ds-DNA 
chain through the nanopore, while bias across the device is fixed. The zgnr 
used for this case has double width than that is used for Fig.[5]. We record 
the characteristic current output for the bases as they translocate through 
the nanopore. Current output is greater than Fig.~[5] but the variation in 
the responses for different basepairs including mismatches decreased slightly.} 

\label{fig7}
\end{figure}

We also check the sensitivity of the device on the nanoribbon width. To investigate this 
we make the zgnr width double than previous results but keep the pore size fixed. In 
Fig.~[7] we plot the sequential determination {\it i.e.}, stop and go translocation of a 
ds-DNA chain containing mismatches through the zgnr-nanopore with increased width. With 
increasing width current output increases, which is trivial as the width increases conductance 
of the device will also increase and so the current. But the sensitivity decreases to some 
extent. As we keep the pore size fixed, the fraction of the current passing around the pore 
will decrease and signature of the basepair will die out with increasing width as the presence 
of the basepairs modify this current only which is detected by the device. For the previous 
case (Fig.~[5]) the range of current variation is 0.09 (arb. unit) for different basepairs 
which reduces to 0.06 (arb. unit) as we doubled the width of the zgnr.

 Following the above results (Fig.~[7]) we can say that there are several issues compete in 
the sequential detection technique. First thing is that to get higher current output from the 
device one has to increase the ribbon width, but it will also hamper device sensitivity to some 
extent. In order to maintain the desired accuracy one has to increase the nanopore dimension 
with increasing ribbon width. Increasing the pore size will increase the fraction of current 
passing around pore and the effect of the basepairs will also become more vivid. Because only 
the changes in the current passing around the nanopore due to the presence of the basepair 
is detected by the device. And to reduce the fluctuations of the basepairs inside the nanopore 
during translocation the edge atoms of the nanopore has to be functionalized with different 
groups~\cite{saha,jeong,ralph} as discussed in the earlier section. Thus, in case of sequential 
determination process of DNA or biomolecules there are several parameters which have to be optimized 
accordingly for accurate and precise measurement.

\end{document}